\documentclass{xjkas}

\def\year{2016} 
\def\month{Feb} 
\def\volume{00} 
\def\issue{0} 
\def\beginpage{1} 
\def\endpage{99} 
\def\received{August 31, 2015} 
\def\accepted{January 10, 2016} 
\date{Received \received; accepted \accepted}

\title{
The Infrared Medium-Deep Survey. V. \\
A New Selection Strategy for Quasars at z $>$ 5 \\
based on Medium-Band Observation with SQUEAN
}

\author[1]{Yiseul Jeon}
\author[1]{Myungshin Im}
\author[2]{Soojong Pak}
\author[1]{Minhee Hyun}
\author[2]{Sanghyuk Kim}
\author[1]{Yongjung Kim}
\author[2]{Hye-In Lee}
\author[2]{Woojin Park}

\affil[1]{Center for the Exploration of the Origin of the Universe (CEOU), Astronomy Program, Department of Physics \& Astronomy, Seoul National University, 1 Gwanak-ro, Gwanak-gu, Seoul 151-742 Korea; \email{ysjeon@astro.snu.ac.kr, mim@astro.snu.ac.kr}}
\affil[2]{School of Space Research and Institute of Natural Sciences, Kyung Hee University, 1732 Deogyeong-daero, Giheung-gu, Yongin-si, Gyeonggi-do 446-701, Korea}

\def\corrauthor{
M. Im
}

\def\runningauthor{
Y. Jeon
}

\def\runningtitle{
IMS VII. Medium-Band Observation with SQUEAN
}

\def\keywords{
observations -- quasar: general -- quasar: supermassive black holes -- surveys
}

\def\abstracttext{
Multiple color selection techniques have been successful in identifying quasars from wide-field broad-band imaging survey data.
Among the quasars that have been discovered so far, however, there is a redshift gap at $5 \lesssim {\rm z} \lesssim 5.7$ due to the limitations of filter sets in previous studies. 
In this work, we present a new selection technique of high redshift quasars using a sequence of medium-band filters: nine filters with central wavelengths from 625 to 1025 nm and bandwidths of 50 nm. Photometry with these medium-bands traces the spectral energy distribution (SED) of a source, similar to spectroscopy with resolution R $\sim$ 15.
By conducting medium-band observations of high redshift quasars at 4.7 $\leq$ z $\leq$ 6.0 and brown dwarfs 
(the main contaminants in high redshift quasar selection) using 
the SED camera for QUasars in EArly uNiverse (SQUEAN) on the 2.1-m telescope at the McDonald Observatory, we show that these medium-band filters are superior to multi-color broad-band color section in separating high redshift quasars from brown dwarfs. 
In addition, we show
that redshifts of high redshift quasars can be determined to an accuracy of $\Delta{\rm z}/(1+{\rm z}) = 0.002$ -- $0.026$. The selection technique can be extended to z $\sim$ 7, suggesting that the
medium-band observation can be powerful in identifying quasars even at the re-ionization epoch.
}

\begin{document}
\jkashead 


\section{Introduction}
Recently, more than 0.4 million quasars have been discovered \citep{fles15}, among which 70 lie at z $\gtrsim$ 6
\citep[e.g.,][]{fan03, fan06131, will07, will10, jian08, jian09, jian15, mort09, mort11}.
Using these high redshift quasars, we can study the evolution of supermassive black holes 
\citep[SMBHs; e.g.,][]{jian07, kurk07, jun15}
and the metal enrichment history of the early universe when the age of the universe was less than 1 Gyr 
\citep[e.g.,][]{jian07, kurk07, juar09, dero11, dero14}. 
In addition, they allow us to examine the line-of-sight condition of the intergalactic medium (IGM), such as the re-ionization state
\citep[e.g.,][]{fan06132, cari10, mcgr11}.



However, there exist redshift gaps in current high redshift quasar surveys, especially in the range of 4.8 $\leq$ z $\leq$ 5.7. 
Although some quasar surveys were performed at z $\sim$ 5 or z $\sim$ 6 
\citep[e.g.,][]{zhen00, shar01, schn01, fan03, fan06131, maha05, cool06, will07, will10, jian08, jian09, jian15, wu10, matu13, mcgr13},
they could not detect quasars at $5 \lesssim {\rm z} \lesssim 5.7$ efficiently 
\citep[Figure 16 in][]{jun15}.
This was due to the limitations of the filter systems currently employed by these surveys as described below.  

Most of the quasars at z $\sim$ 5 were discovered by the Sloan Digital Sky Survey \citep[SDSS;][]{rich02, mcgr13}, 
using the $r-i$ color for selecting $r$-dropouts and the $i-z$ color for removing brown dwarfs. 
Also \citet{iked12} used the $r-i$ vs. $i-z$ color-color diagram and discovered only one type-2 quasar at ${\rm z}=5.07$. 
However, these studies were unsuccessful in finding quasars at $5 \lesssim {\rm z} \lesssim 5.7$ because the $r$-dropout quasars, starting from z $\sim$ 4.6, 
 blend  with brown dwarfs on color-color diagrams for z $\sim$ 5.1 (Jeon et al. in preparation). 
\citet{wu10} adopted new color cuts, $J-K$ vs. $i-Y$ 
to find high redshift quasar candidates, but their method is limited to ${\rm z} < 5.3$.
On the other hand, for quasar surveys at z $\sim$ 6, $i-z$ vs. $z-J$ color-color diagrams are usually used. However the $i$-dropouts start from z $\sim$ 5.7; thus, most surveys for z $\sim$ 6 are limited to z $>$ 5.7. Therefore we need a new method
and a new filter system which exploits the wavelength range between conventional filters to find quasars at redshift between 5.1 and 5.7. 

To discover quasars at $5 \lesssim {\rm z} \lesssim 5.7$, we have been conducting an imaging survey as a part of the Infrared Medium-deep Survey (IMS; Im et al. in preparation). This imaging
survey has used
the Camera for QUasars in EArly uNiverse \citep[CQUEAN;][]{park12, kim11, lim13}, with custom-designed $is$ and $iz$ filters that compensate for previous filter systems. Since the central wavelengths of these filters are located between $r$ and $i$, and between $i$ and $z$, respectively, it was expected that these filters will be effective in selecting high redshift quasars at 5 $<$ z $<$ 6, where SDSS or other filter systems cannot explore (Jeon et al. in preparation). The quasar candidates were selected from $r$-dropout objects of $z$ $<$ 19.5 mag over 3,400 deg$^2$ using multi-wavelength data such as SDSS, the United Kingdom Infrared Telescope Infrared Deep Sky Survey \citep[UKIDSS;][]{lawr07} Large Area Survey (LAS), the Two Micron All Sky Survey (2MASS), and the Wide-field Infrared Survey Explorer ({\it WISE}). 
So far, 6 $r$-dropouts in 4.7 $\leq$ z $\leq$ 5.4, and 2 $i$-dropouts in 5.9 $\leq$ z $\leq$ 6.1 have been confirmed as quasars with optical spectroscopic observations (Jeon et al. in preparation); these are referred to as IMS quasars.  
The discovery of 6 $r$-dropout quasars at 4.7 $\leq$ z $\leq$ 5.4 
is encouraging, but the same technique has not been successful at uncovering quasars at 5.4 $<$ z $<$ 5.7.
This is mostly because
the adopted color cuts cannot yet remove contaminants such as late type stars or brown dwarfs
effectively. 
Therefore, 
it is highly desirable to have a new method based on a new filter system for selecting quasars in this range.

\citet{matu13} adopted a spectral energy distribution (SED) fitting method and discovered a faint quasar at z = 5.41 from the Advance Large Homogeneous Area Medium Band Redshift
Astronomical \citep[ALHAMBRA;][]{mole08} survey over $\sim$1 deg$^2$. Their survey included a filter set composed of 20 optical medium-band and 3 broad-band $J/H/Ks$ filters, and an extensive library of SEDs including those of active galaxies. 
Despite the fact that high redshift quasar candidates are contaminated by many low redshift galaxies and late type stars, they were able to select the quasar at z = 5.41 via SED fitting. 
This demonstrates that low resolution spectroscopy with medium-band filters can be an effective way to select high redshift quasars.
  
As an attempt to improve the selection of high redshift quasars, we devised a set of 9 medium-band filters to better trace SEDs of various sources and 
modified CQUEAN to allow the usage as many as 20 filters. 
We now call this modified CQUEAN the SED 
camera for QUasars in EArly uNiverse \citep[SQUEAN; Kim et al. in preparation;][]{choi15}.
Like CQUEAN, SQUEAN operates on the 2.1 m Otto Struve Telescope at the McDonald Observatory. 
To explore the potential of selecting high redshift quasars using our medium-band filters,
we observed known high redshift quasars and brown dwarfs, and 
conducted SED fitting to distinguish these two populations. 
This paper reports the result of the study.
Section \ref{c_obser} describes the overview of our new medium-band filters and two test observations. In Sections \ref{c_prop} and \ref{c_fit}, we describe the performance from the result of medium-band photometry including SED fitting results. We discuss the
feasibility of using medium-band filters on small or medium-class telescopes for future high redshift quasar selection
in Section \ref{c_stra}.
We summarize our results in the final section. Throughout this paper, we use a cosmology with $\Omega_M=0.3, \Omega_\Lambda=0.7,$ and $H_0$ = 70 km s$^{-1}$Mpc$^{-1}$, and the AB magnitude system.


\section{Observations} \label{c_obser}

\subsection{SQUEAN Medium-band Filters} \label{filter}
The SQUEAN medium-band filter set consists of nine filters, $m625$, $m675$, $m725$, $m775$, $m825$, $m875$, $m925s$, $m975$, and $m1025$, 
where ''m'' stands for ''medium'' and the number indicates the central wavelength in nm
(Table \ref{tbl_filter}).
The bandwidths of these filters are fixed to 50 nm. 
In addition, SQUEAN has an $m925n$ filter, whose bandwidth is 25 nm and has a central wavelength 
identical to that of $m925s$. 
These filters are off-the-shelf products from Edmund Optics, Inc..
The optical characteristics of the filters are described in a separated SQUEAN instrument paper (Kim et al. in preparation).

\begin{figure}
\centering
\includegraphics[scale=.34]{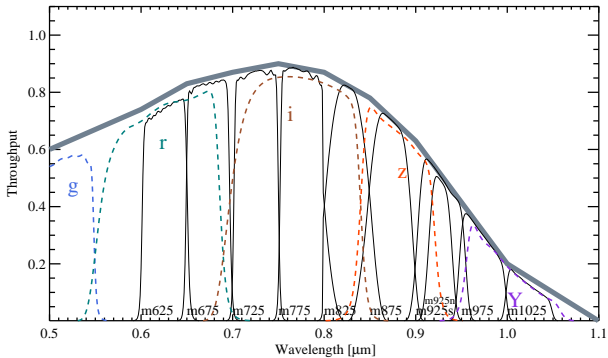}
\caption{The QE of SQUEAN (thick solid line) and the combined response function of the detector QE and filters. The thin solid lines are for our medium-band filters and the dashed lines are for the CQUEAN $g/r/i/z/Y$. \label{fig_filter}}
\end{figure}

Figure \ref{fig_filter} shows filter transmission curves of the medium-band filters (thin solid lines) and other SQUEAN filters (dashed lines) with the detector quantum efficiency (QE; thick solid line) taken into account. 
The figure shows that the medium-band filter bandwidths are several times narrower than those of broad-band filters, such as SDSS $r/i/z$.



\begin{table}[t!]
\caption{Medium-band filter characteristics\label{tbl_filter}}
\tabcolsep=0.09cm
\centering
\begin{tabular}{lccc}
\toprule
Filter  & Central                       & Effective                     &FWHM$^{\rm a}$   \\
Name    & Wavelength (nm)       &Wavelength$^{\rm a}$ (nm)        &(nm)\\
\midrule
$m625$&    625&  627&   50\\
$m675$&    675&  673&   48\\
$m725$&    725&  726&   49\\
$m775$&    775&  777&   51\\
$m825$&    825&  827&   47\\
$m875$&    875&  870&   49\\
$m925s$&    925&  926&   52\\
$m925n$&    925&  926&   26\\
$m975$&    975&  969&   49\\
$m1025$&   1025& 1018&   47\\
\bottomrule
\end{tabular}
\tabnote{
Note. 
\\ $^{\rm a}$  Kim et al. in preparation
}
\end{table}

\subsection{SQUEAN Medium-band Observations} \label{obser}
We carried out test observations 
with a subset of medium-band filters installed in CQUEAN
in 2014 November and 
with a complete set of medium-band filters in SQUEAN in
2015 February.
During the two observing runs, we observed 6 IMS quasars at 4.7 $\leq$ z $\leq$ 6.0 and 5 brown dwarfs (Table \ref{tbl_sample}) with $m675$, $m725$, $m775$, $m825$ and $m875$ filters. The observations were carried out under mostly clear sky conditions. 
However, both observations were executed during bright time, and seeing values varied from 1.3'' to 4.0''.

Preprocessing, such as bias subtraction, dark subtraction and flat fielding were conducted with  usual data reduction procedures using the IRAF\footnote{IRAF is distributed by the National Optical Astronomy Observatory, which is operated by the Association of Universities for Research in Astronomy,  Inc.,  under cooperative agreement with the National Science Foundation.}   
\texttt{noao.imred.ccdred} package. Then we combined images of each field and each filter in average. We used the \texttt{ccmap} task of IRAF and SCAMP \citep{bert06} to derive astrometric solutions. 

SExtractor \citep{bert96} was used for source detection and photometry. We derived auto-magnitudes which are taken as the total magnitudes. 
We estimated the limiting magnitudes for each medium-band filter using the data taken on 2015 February 8. 
The 5$\sigma$ detection limits of a point source with 1 hour integration time
are 23.7 mag for $m675$ and $m725$, 23.3 mag for $m775$ and $m825$, and 22.7 mag for $m875$-bands.

\subsection{Photometric Calibration} \label{calib}
\begin{figure}
\centering
\includegraphics[scale=.41]{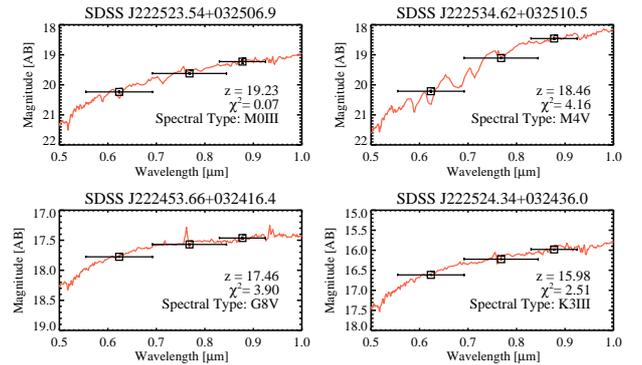}
\caption{Examples of the $\chi^2$ fitting of SDSS stars. The solid line shows the best-fit template and the boxes are from SDSS $r/i/z$ photometry. 
\label{fig_sedst}}
\end{figure}

\begin{figure}
\centering
\includegraphics[scale=.46]{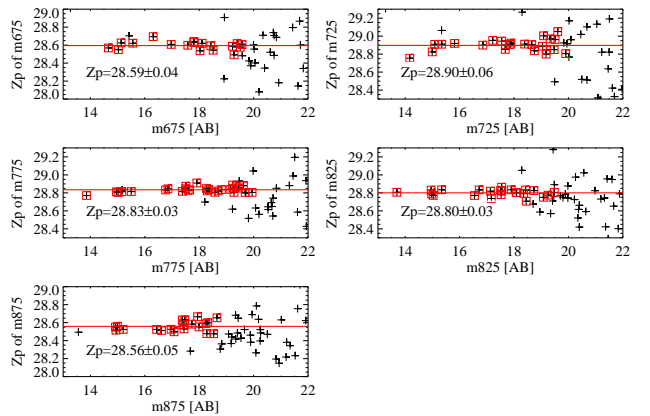}
\caption{The $Zp$ calculations of IMS J143704.82+070808.3 field for each filter. The crosses are for all sources in the field and the boxes are sources that satisfy the conditions described in Section \ref{calib}.\label{fig_calzp}}
\end{figure}

For photometric calibration, we used stars that have SDSS photometry and are in the same field as our target. Using SDSS $r/i/z$ magnitudes of a star, we did a $\chi^2$ fitting to the SED of the object to find the best spectral template among stellar spectral templates. We used the stellar templates from \citet{gunn83}, that contain 175 spectra with various stellar types. From the best-fit template, we calculated model magnitudes for each medium-band filter by convolving their filter transmission curves and defined these values as their standard magnitudes
($M_{\rm std}$) of each filter. Figure \ref{fig_sedst} shows examples of the $\chi^2$ fitting of stars, with their $r/i/z$ magnitudes (boxes) and the best-fit templates (solid lines). Then, the $M_{\rm std}$  of the stellar sources were compared with their instrument magnitudes ($m_{\rm ins}$; $m_{\rm ins}=-2.5\log({\rm DN})$), and the differences between the $M_{\rm std}$ and the $m_{\rm ins}$ were used to estimate the zero-point ($Zp$). During the $Zp$ calculation, faint objects or objects with large reduced $\chi^2$ ($\chi^2_{\nu}$) values were rejected: we chose stars of $\chi^2_{\nu}$ $<$ 5 and instrument magnitude error $<$ 0.05 mag for each band. Figure \ref{fig_calzp} shows an example of the $Zp$ calculations for each filter in the case of IMS J143704.82+070808.3. The crosses are for all sources in the field and the boxes are for sources that satisfy the conditions explained above. After 3$\sigma$ clipping, we calculated the $Zp$ as the average value of the magnitude differences. For the $Zp$ uncertainty, we used the standard deviation value of the $Zp$ values. The average zero-point error is about 0.05 mag. The standard magnitudes of our targets were calculated as $M_{\rm std}=Zp+m_{\rm ins}$. 

This calibration technique was compared to results from standard star observations. We used data from the observation of 2015 February 5. BD+33 2642 
was observed with the 5 filters at two different airmasses. To calculate the standard magnitudes of the star for each filter, we used its spectrum obtained from \citet{oke90}, and the filter transmission curves. 
The $Zp$ values from the standard star data are found to be consistent with the values derived from SDSS stars.

\section{Properties of Medium-band Photometry} \label{c_prop}

\subsection{Filter Response Curves with Quasar Spectra} \label{qsed}
\begin{figure}
\centering
\includegraphics[scale=.43]{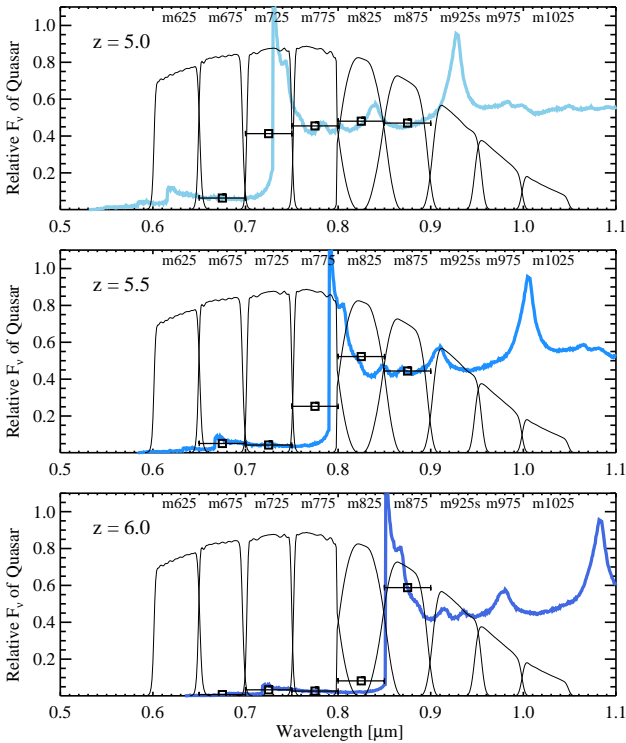}
\caption{Filter response curves (thin solid lines) with quasar spectra (thick solid lines) 
redshifted to z = 5.0, 5.5, and 6.0. 
The model magnitudes in each filter (squares) trace the SEDs of the sources similar to spectroscopy with R $\sim$ 15.
\label{fig_qsosed}}
\end{figure}

We plot the filter response curves of medium-band filters (thin solid lines) with quasar SED templates at different redshifts (thick solid lines) in Figure \ref{fig_qsosed}. For the quasar SEDs, we used the median composite spectrum of SDSS quasars from \citet{vand01} and adopted the IGM attenuation model from \citet{mada96}. We also plot the model magnitudes from the SED templates in each filter with squares. We can notice that the Lyman $\alpha$ (Ly$\alpha$) emission line is located in the $m725$-band at z = 5.0, in the $m775$-band at z = 5.5 and between the $m825$ and $m875$-bands at z = 6.0, and the model magnitudes show a sharp drop in flux shortward of these filters. We can see that the photometry with these medium-bands trace the SED of sources, similar to spectroscopy with  resolution R $\sim$ 15.

\subsection{Medium-band Spectral Energy Distributions} \label{mags}
Table \ref{tbl_mag} lists the medium-band magnitudes of our targets with $g/r/i/z/Y/J/H/K$-band photometry from SDSS and UKIDSS LAS.
The magnitude errors of the medium-band filters are computed by combining the standard SExtractor estimates based on Poisson statistics and the zero-point errors from Section \ref{calib}.  



\begin{figure*}
\centering
\includegraphics[scale=.75]{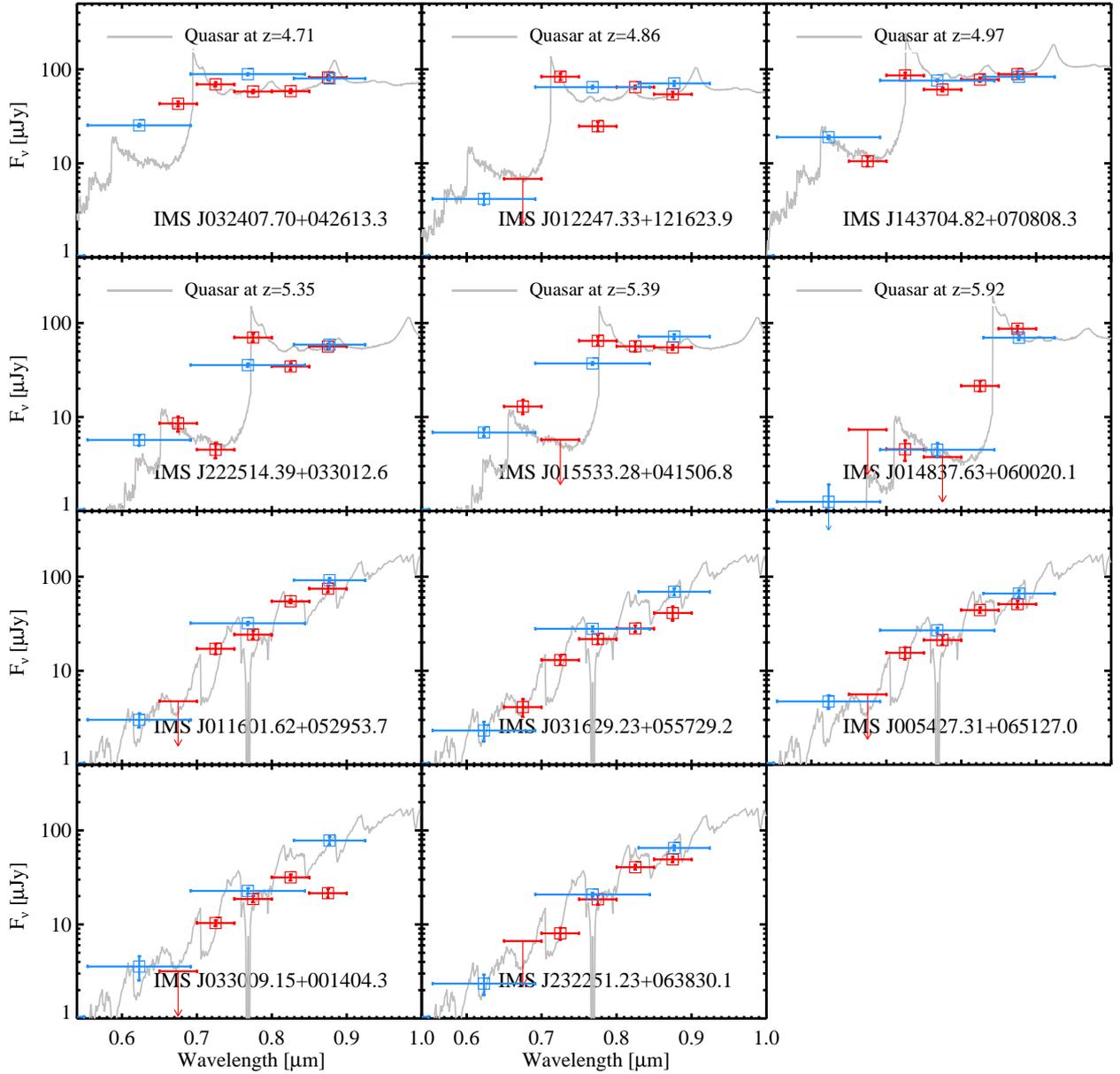}
\caption{Medium-band SEDs of quasars and brown dwarfs from SQUEAN observations (red squares). 
Also we plot the SDSS $r/i/z$ data (blue squares), the redshifted quasar SEDs (gray lines in upper two rows), and the model brown dwarf spectra (gray lines in lower two rows).
This figure demonstrates that high redshift quasar SEDs and brown dwarf SEDs can be distinguished easily thanks to the fine
wavelength sampling of the medium-band filters.
 \label{fig_sedall}}
\end{figure*}

The medium-band filters trace SEDs of high redshift quasars and brown dwarfs differently.
In Figure \ref{fig_sedall}, we show the medium-band SEDs of quasars and brown dwarfs, where the red squares are the medium-band data and the blue squares are the SDSS $r/i/z$ magnitudes. 
The gray lines in the upper two rows are quasar templates from \citet{vand01} redshifted appropriately, with IGM attenuation \citep{mada96}. The gray lines in the lower two rows are model brown dwarf spectra with a temperature of 2,200K \citet{burr06}.
Note that IMS J012247.33+121623.9 is a broad absorption line (BAL) quasar, which shows a deep absorption feature near the wavelengths of the $m775$-band. 
High redshift quasars show 
a sharp drop in flux in the filter that covers 
the wavelength range blueward from 
the redshifted Ly$\alpha$ emission line.
On the other hand, 
SEDs of brown dwarfs smoothly increases from short to long wavelengths.
Such a difference in SEDs is much less pronounced in the
$r/i/z$ broad-band photometry. The drastic magnitude drop at the redshifted Ly$\alpha$ is  key to distinguishing high redshift quasars from brown dwarfs, and the medium-band filters 
offer sufficient spectral resolution for such a purpose. These dropouts can also be recognized in color-color diagrams. 

\begin{figure}[t]
\centering
\includegraphics[scale=.45]{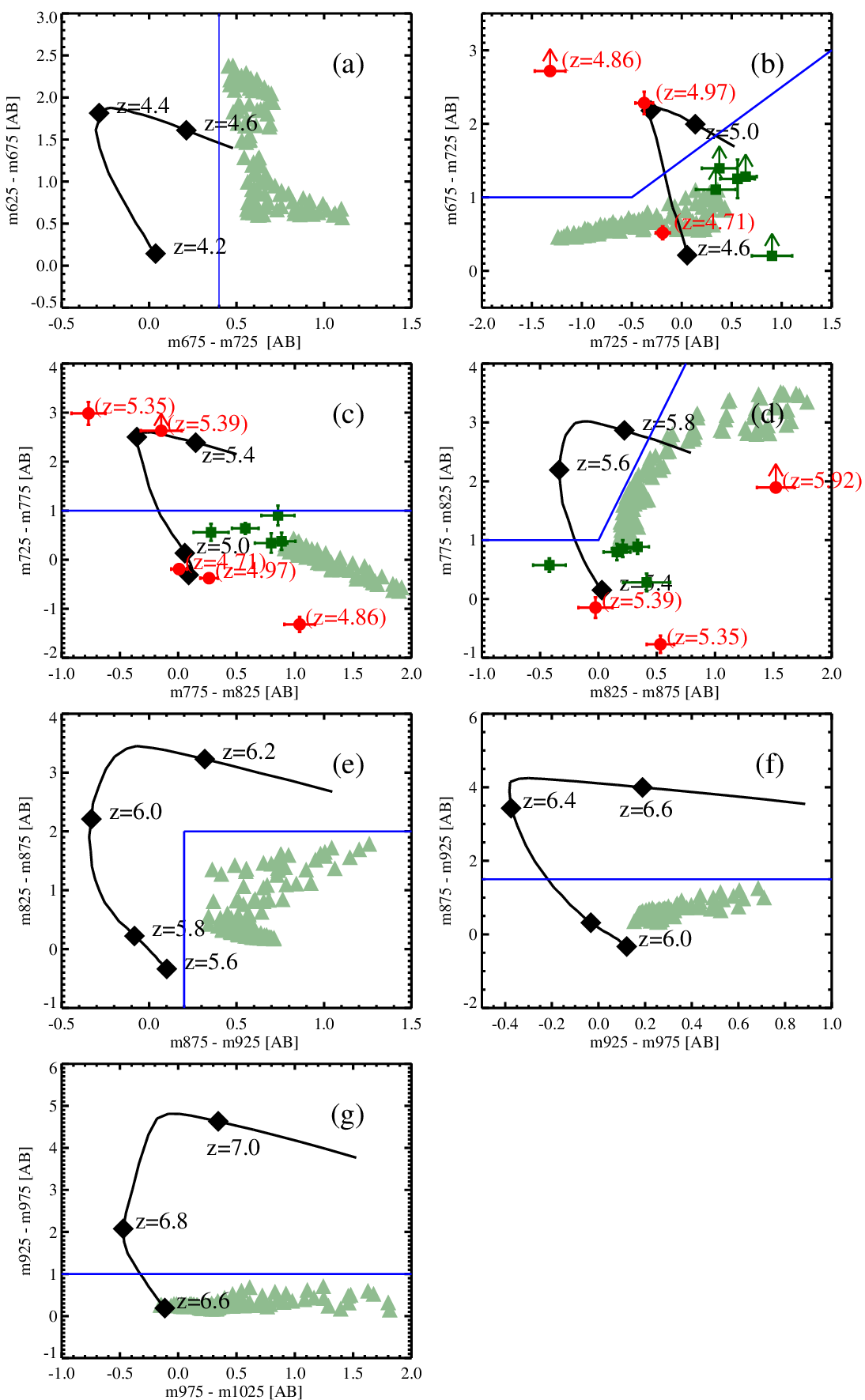}
\caption{Color-color diagrams of successive combinations of nine medium-band filters. 
The black lines are quasar redshift tracks with diamond symbols indicating redshifts at a $\Delta {\rm z} = 0.2$ interval,
except the last redshift point is chosen to be the redshift where Ly$\alpha$ is located at 12.5 nm above the central wavelength of the second filter.
The light green triangles are brown dwarfs, 
the dark green boxes are IMS brown dwarfs, 
and the red circles are IMS quasars.
\label{fig_ccdall}}
\end{figure}

\subsection{Color-color Diagrams} \label{ccds}
Figure \ref{fig_ccdall} shows 7 color-color diagrams from combinations of all nine medium-band filters (black lines are quasar redshift tracks within specific redshift ranges, light green triangles are model brown dwarfs from \citet{burr06}, dark green boxes with error bars or arrows are IMS brown dwarfs, and red circles with error bars or arrows are IMS quasars). Figure \ref{fig_ccdall} (a) is for $m625 - m675$, vs. $m675- m725$, (b) is for $m675 - m725$, vs. $m725 - m775$, (c) is for $m725 - m775$ vs. $m775 - m825$, (d) is for $m775 - m825$ vs. $m825 - m875$, (e) is for $m825 - m875$ vs. $m875 - m925$, (f) is for $m875 - m925$ vs. $m925 - m975$, and (g) is for $m925 - m975$ vs. $m975 - m1025$. Figure \ref{fig_ccdall} (b), (c), and (d) contain the medium-band colors from our observations. From Figure \ref{fig_ccdall} (b), we can deduce that the $m675$-dropout is found for IMS J012247.33+121623.9 (z=4.86) and IMS J143704.82+070808.3 (z=4.97). 
On the other hand, it is difficult to select IMS J032407.70+042613.3 (z=4.71) as a high redshift quasar, because its Ly$\alpha$ drop is located at the $m675$-band and its colors using these 3 filters are similar to those of brown dwarfs. 
Note that the BAL quasar at z=4.86, IMS J012247.33+121623.9, is off from the redshift tracks due to broad absorption lines in its continuum.  
For similar reasons, in Figure \ref{fig_ccdall} (c), IMS J222514.39+033012.6 (z=5.35) and IMS J015533.28+041506.8 (z=5.39) also show red $m725 - m775$ colors relative to IMS J032407.70+042613.3 (z=4.71), IMS J012247.33+121623.9 (z=4.86), and IMS J143704.82+070808.3 (z=4.97) due to the $m725$-dropout. 
We can see that Figure \ref{fig_ccdall} (b), (c), and (d) can be used for selecting quasars at 4.8 $<$ z $<$ 5.8 because the  positions of the quasars are  significantly different from those of the brown dwarfs, distanced by $\gtrsim$ 1 mag from the quasar tracks. 
For the quasar selection at z $>$ 5.8, Figure \ref{fig_ccdall} (e), (f), and (g) show that the quasars at z $>$ 5.8 can be separated from brown dwarfs clearly.
Therefore we can use these color-color diagrams from medium-band photometry for quasar candidate selections at 4.7 $<$ z $<$ 7.
Table \ref{tbl_zrange} lists the color cuts (blue solid lines in Figure \ref{fig_ccdall}) and the redshift ranges for quasar selection.
Since Lyman break galaxies (LBGs) at high redshift 
have ultraviolet SEDs similar to those of quasars, these selection criteria can also be used for selecting 
high redshift star-forming galaxies. 
On the other hand, we expect that the objects satisfying the color cuts at bright magnitude \citep[$< 21$ AB mag; e.g.,][]{shim07} are predominantly quasars.

\begin{table}[t!]
\caption{Color cuts and redshift ranges for quasar selection \label{tbl_zrange}}
\tabcolsep=0.09cm
\centering
\begin{tabular}{clcc}
\toprule
Figure & Selection criteria& Redshift range \\
\midrule
(a) & $m675 - m725 < 0.4$&4.2 $<$ z $<$ 4.6\\
(b) & $m675 - m725 > 1$ &4.7 $<$ z $<$ 5.1\\
    & $\cap~~ m675 - m725 > m725 - m775 +1.5$ & \\
(c) & $m725 - m775 > 1$&5.1 $<$ z $<$ 5.5\\
(d) & $m775 - m825 > 1$&5.5 $<$ z $<$ 5.8\\
    & $\cap~~ m775 - m825 > 4\times(m825 - m875) +1$& \\
(e) & $m825 - m875 > 2$&5.8 $<$ z $<$ 6.3\\
    & $\cup~~ m875 - m925 < 0.2$&\\
(f) & $m875 - m925 > 1.5$&6.3 $<$ z $<$ 6.7\\
(g) & $m925 - m975 > 1$&6.7 $<$ z $<$ 7.1\\
\bottomrule
\end{tabular}
\end{table}

\begin{figure}
\centering
\includegraphics[scale=.45]{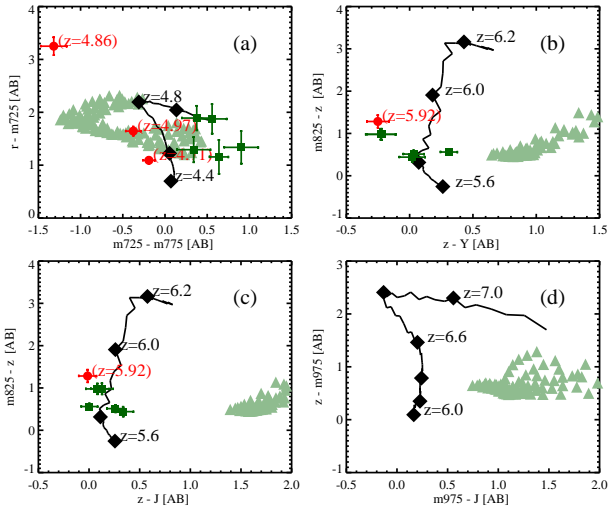}
\caption{Similar to Figure \ref{fig_ccdall}, but including SDSS and UKIDSS LAS filters. 
\label{fig_ccdsds}}
\end{figure}

For comparison, we plot various color-color diagrams using SDSS and UKIDSS LAS filters in Figure \ref{fig_ccdsds}. 
Since the $r$-band has wavelengths similar to those of $m675$, we plot the $r - m725$ vs. $m725 - m775$ color-color diagram in Figure \ref{fig_ccdsds} (a). 
In Figure \ref{fig_ccdsds} (a), 
the colors of the model quasar redshifted to 4.7 $<$ z $<$ 5.1 are not different from those of brown dwarfs, contrary to Figure \ref{fig_ccdall} (b).
This is due to the difference in wavelength coverage between the $r$ and $m675$-bands. 
Since the $m675$ filter has a narrower bandwidth compared to the $r$ filter, the $m675-m725$ color can measure the Ly$\alpha$ dropout feature clearly, without being affected by other spectral features. For example, IMS J143704.82+070808.3 at z = 4.97 shows a higher flux in the $r$-band than in the $m675$-band because the $r$ filter also covers the Lyman $\beta$ emission line at this redshift.
Therefore, the dropout effect between $r$ and $m725$ is not strong enough to distinguish quasars from brown dwarfs. 

To check the possibility of selecting quasars at z $\sim$ 6 with our medium-bands and near-infrared bands, we adopted the $Y$ or $J$-bands ((b) and (c) of Figure \ref{fig_ccdsds}). The $z - Y$ or $z - J$ colors of z $\sim$ 6 quasars are different from those of brown dwarfs, which is used in a traditional selection method to select z $\sim$ 6 quasars, such as the $i - z$ vs. $z - J$ color-color diagram
\citep[e.g.,][]{fan01, fan03, fan04, fan06131, jian08, jian09, will07, will09, will10}. 
Therefore, using these three diagrams, we are able to select quasars at z $\sim$ 6 that are separated from the brown dwarfs. 

In Figure \ref{fig_ccdsds} (d), the $z-m975$ vs. $m975-J$ color-color diagram shows a clear separation between the quasar redshift track and brown dwarfs, similar to Figure \ref{fig_ccdall} (g), which is also suitable for selecting z $\sim$ 7 quasar candidates.

\section{SED Fitting from Medium-band Photometry}\label{c_fit}
To estimate the efficiency of the medium-band filters, we performed SED fitting using templates of the SDSS quasar composite spectrum from \citet{vand01} with IGM attenuation \citep{mada96}. 
SEDs from the medium-bands or other bands are compared to the quasar templates redshifted to various redshifts and $\chi^2_{\nu}$ values between the SEDs and the quasar templates were calculated. 
We found the best-fit template of a source using the minimization of $\chi^2_{\nu}$ values and estimated a photometric redshift (z$_{\rm phot}$) from the template.

We used several filter sets for the SED fitting, combining our medium-bands and SDSS filters, to compare fitting performances while using different filter sets. We define the filter sets as follows:\\
\textsf{Filterset1}=$\{m675, m725, m775, m825, m875\}$\\
\textsf{Filterset2}=$\{r, m725, m775, m825, m875\}$\\
\textsf{Filterset3}=$\{m675, m725, m775, m825, z\}$\\
\textsf{Filterset4}=$\{r, m725, m775, m825, z\}$\\
\textsf{Filterset5}=$\{r,i,z\}$


\subsection{Best-fit $\chi^2_{\nu}$ Values of Quasars and Brown Dwarfs} \label{chi_1}

\begin{figure*}
\centering
\includegraphics[scale=.7]{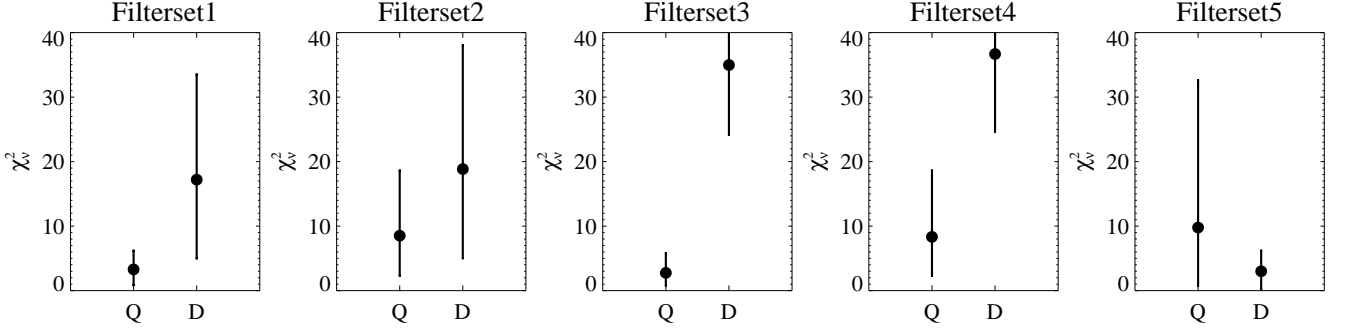}
\caption{
The average values of the best-fit $\chi^2_{\nu}$ values for quasars and brown dwarfs from each filter set. We exclude IMS J012247.33+121623.9, since it is a BAL quasar.
The ranges of the best-fit $\chi^2_{\nu}$ values are denoted with error bars. 
\label{fig_sdssz}}
\end{figure*}

After finding the best-fit templates for each target, we compared the best-fit $\chi^2_{\nu}$ values for quasars and brown dwarfs for a specific filter set. 
Figure \ref{fig_sdssz} shows the average values of the best-fit $\chi^2_{\nu}$ values for quasars and brown dwarfs for each filter set. 
We exclude IMS J012247.33+121623.9, since it is a BAL quasar, whose spectra are known to be clearly different from normal quasar spectra.
The ranges of the best-fit $\chi^2_{\nu}$ values are denoted with error bars. 
We can see that the best-fit $\chi^2_{\nu}$ values from the quasar and the brown dwarfs are quite  different, although some brown dwarfs show small $\chi^2_{\nu}$ values comparable to those of quasars. 
On average, for \textsf{Filterset1 -- 4}, which include 3 medium-bands that substitute the $i$-band wavelengths ($m725, m775, m825$), the best-fit $\chi^2_{\nu}$ values for the quasars are  2 -- 13 times smaller than those for the brown dwarfs. 
Therefore the SED fitting from photometry using SDSS filters and these 3 medium-band filters can be used for quasar candidate selection at 5 $<$ z $<$ 6 efficiently:
we can choose robust quasar candidates among large samples selected from broad-band photometry, 
rejecting contaminants that have large $\chi^2_{\nu}$ values. 
On the other hand, in the case of \textsf{Filterset5}, which contains only $r/i/z$-bands without medium-bands, the best-fit $\chi^2_{\nu}$ values of the brown dwarfs are instead smaller than those of the quasars. Therefore we might misclassify brown dwarfs as actual quasars via SED fitting if we use photometry from the $r/i/z$-bands only.
We tested with another filter set $\{r, m725, m775, z\}$ and found that this filter set shows poor ability to separate quasars from brown dwarfs using their best-fit $\chi^2_{\nu}$ values.


\subsection{Photometric Redshifts of Quasars} \label{chi_2}

\begin{figure*}
\centering
\includegraphics[scale=.7]{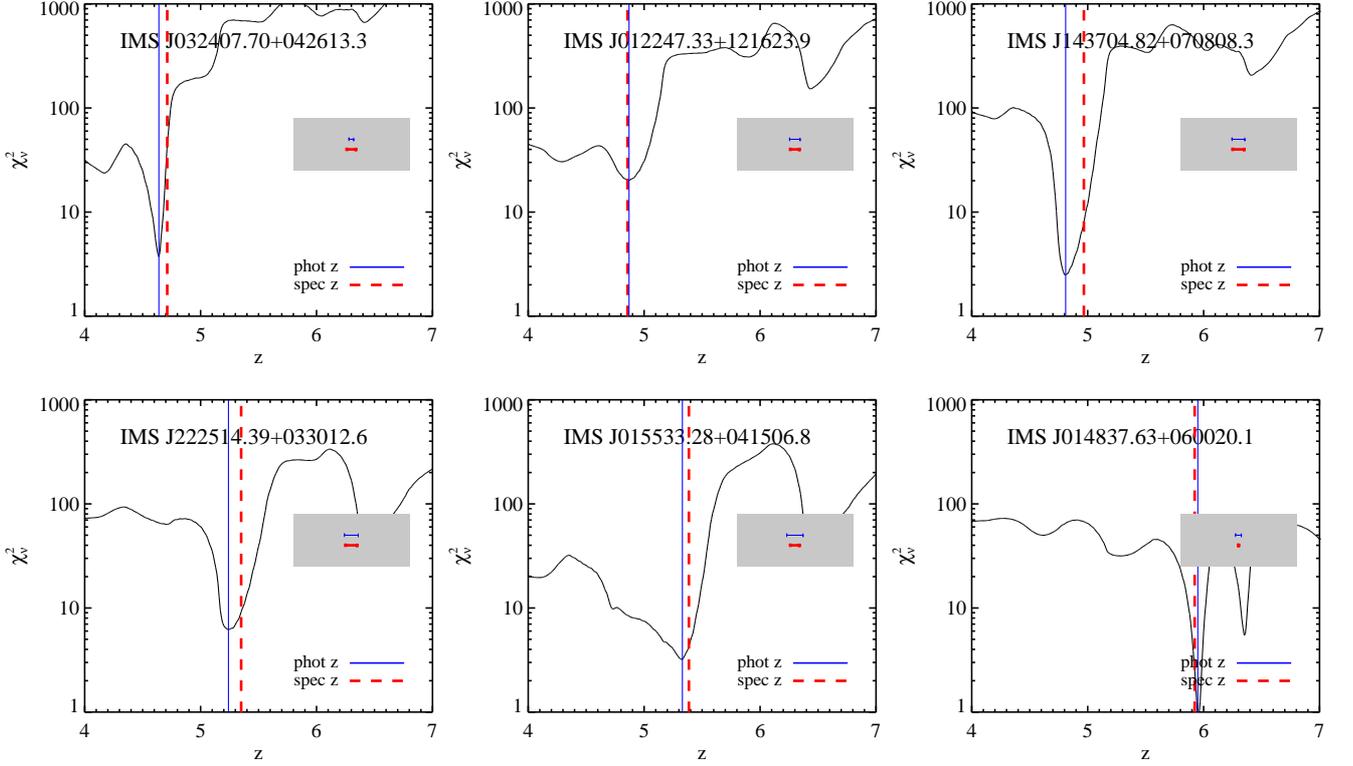}
\caption{The $\chi^2_{\nu}$ values as a function of redshift for each quasar when we use 
 \textsf{Filterset1}.
The red dashed lines indicate the z$_{\rm spec}$ values
 and the blue solid lines indicate for the z$_{\rm phot}$ values. 
The errors of the z$_{\rm phot}$ (blue) and z$_{\rm spec}$ (red) are denoted in the gray boxes.
\label{fig_photz}}
\end{figure*}

\begin{figure}
\centering
\includegraphics[scale=.55]{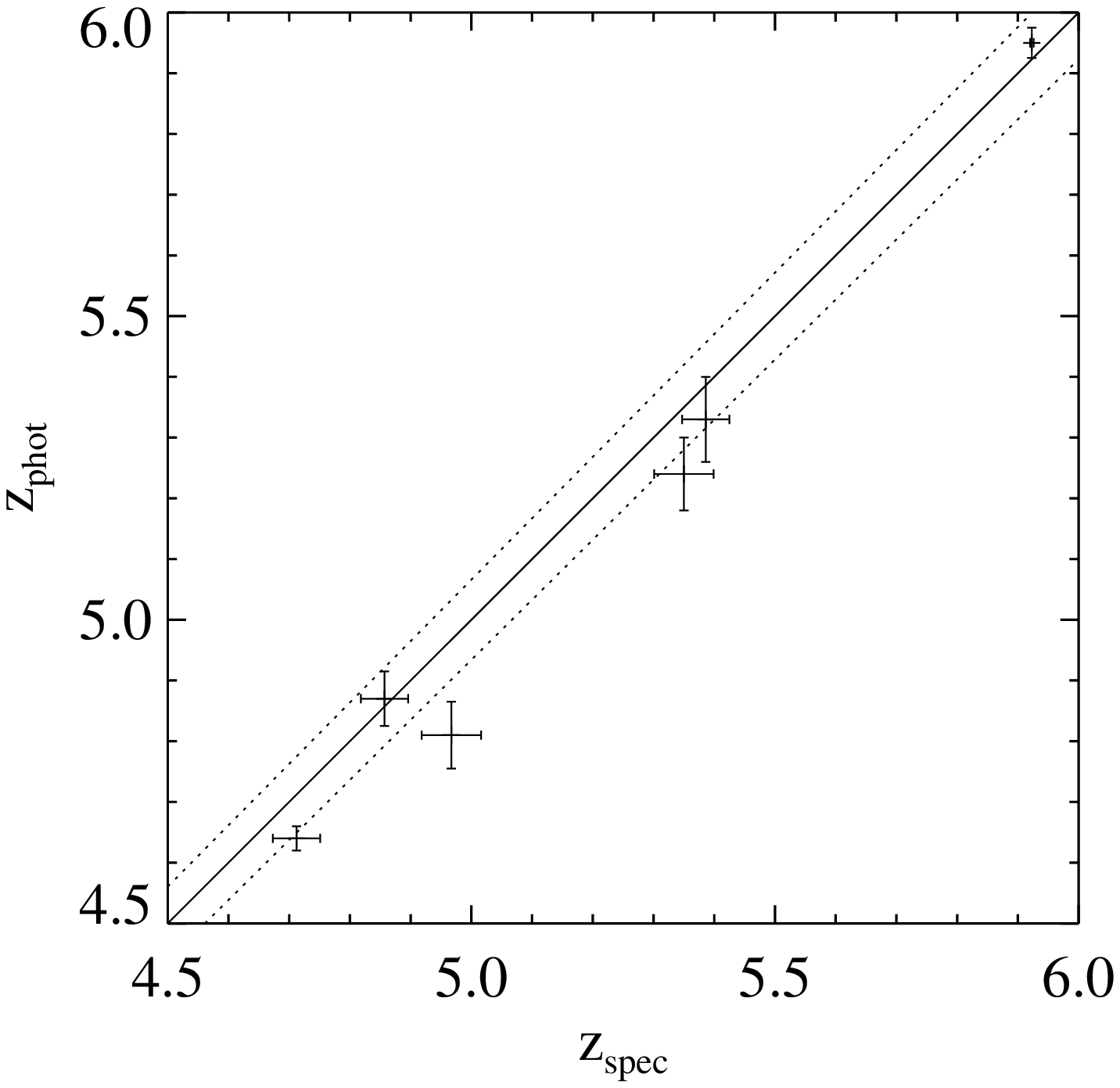}
\caption{
Photometric redshifts from medium-band photometry versus spectroscopic redshifts of high redshift quasars. 
The dotted lines indicate $\Delta{\rm z} /(1 + {\rm z})$ = 0.012.
The solid line indicates the case where z$_{\rm spec}$=z$_{\rm phot}$.
The comparison shows that medium-band data can provide photometric
redshifts with an accuracy of $\sim$1\%.
\label{fig_specz}}
\end{figure}

Using the 5 medium-band filters from $m675$ to $m875$, we estimated z$_{\rm phot}$ via $\chi^2$ fitting. Figure \ref{fig_photz} shows the $\chi^2_{\nu}$ distributions as a function of redshift for each quasar. 
The z$_{\rm phot}$ values are estimated from redshifts where $\chi^2_{\nu}$ has a minimum value. The red dashed lines indicate z$_{\rm spec}$ and the blue solid lines are for z$_{\rm phot}$. 
The errors of the z$_{\rm phot}$ (blue) and z$_{\rm spec}$ (red) are denoted in the gray boxes on each plot. 
The uncertainties of z$_{\rm phot}$ ($\Delta$z$_{\rm phot}$) are estimated from the 1$\sigma$ deviations (confidence level of 68.3\%) of the $\chi^2_{\nu}$ distribution, where the $\chi^2_{\nu}$ takes on a value one greater than its minimum. 
The z$_{\rm spec}$ uncertainties are estimated as discussed in Jeon et al. (in preparation).
We can see that z$_{\rm phot}$ agree with z$_{\rm spec}$ within their redshift uncertainties. 
The $\Delta$z$_{\rm phot}$ values are less than 0.07, which shows the power of the medium-band photometry  to estimate accurate z$_{\rm phot}$ of high redshift quasars via $\chi^2$ fitting of SEDs. The z$_{\rm phot}$ values and their errors are provided in Table \ref{tbl_sample}.
Figure \ref{fig_specz} shows z$_{\rm phot}$ versus z$_{\rm spec}$,
with the solid line indicating z$_{\rm spec}$=z$_{\rm phot}$,
and this figure also demonstrates the excellent agreement between the two values.
The average absolute deviation of z$_{\rm phot}$ $-$ z$_{\rm spec}$ ($\Delta{\rm z}$)
from this comparison is only 0.012 in $\Delta{\rm z} /(1 + {\rm z})$ (dotted lines), which is better than the value we get from the formal error of the fit.

\citet{rich09, rich15} provide z$_{\rm phot}$ values for 3 of the 6 quasars, IMS J012247.33+121623.9, IMS J143704.82+070808.3, and IMS J222514.39+033012.6, estimated from broad-band photometry, $u/g/r/i/z/J/H/K$.
Their estimate for IMS J012247.33+121623.9 (z$_{\rm phot}=5.455^{+0.135}_{-0.095}$ from \citet{rich15}) does not agree with our spectroscopic redshift, 
although IMS J143704.82+070808.3 (z$_{\rm phot}=5.075^{+0.505}_{-0.455}$ from \citet{rich09} or $5.265^{+0.115}_{-0.505}$ from \citet{rich15}) and IMS J222514.39+033012.6 (z$_{\rm phot}=5.415^{+0.285}_{-0.395}$ from \citet{rich15}) are in agreement with our estimates.
Also these estimates show large $\Delta$z$_{\rm phot}$ values compared to those from our medium-band photometry.

\section{Observation Strategy} \label{c_stra}
In this section, we 
discuss whether medium-band filters can be used on small to medium-sized telescopes for high redshift quasar selection.
We showed that three medium-bands ($m725, m775,$ and $m825$ filters) are sufficient to remove brown dwarfs through the SED fitting when SDSS photometric information is provided. Considering the depths of the medium-bands and the average magnitude of our candidates ($i$ = 19.9 mag and $z$ = 19.3 mag), we need 6, 6, and 3 minutes for $m725$, $m775$, and $m825$-band observations with the McDonald 2.1-m telescope and SQUEAN, respectively, to reach a S/N $\sim$ 25 (corresponding photometric error of 0.04 mag) for detection or a 5$\sigma$ upper limit at 6$\mu$Jy for dropouts in $m725$ and $m775$. In this case we need 15 minutes per target for photometry in these three medium-bands. 
If a similar filter set is used on a 0.5-m class telescope \citep[e.g., LSGT;][]{im15}, several hours of observing time is necessary for each target.



It is possible to extend a similar selection method to higher redshifts (z = 6.0, 6.5 and 7.0; see Figures \ref{fig_ccdall} (e), (f), and (g)). 
A particularly interesting point is of the redshift range
at ${\rm z} > 6.4$ where the lack of
wide-field near-infrared imaging survey data and severe brown dwarf contamination make 
quasar selection difficult.
A luminous quasar at ${\rm z}=7.085$ \citep{mort11} has $Y=20.3$ mag and
Ly$\alpha$ at 985 nm. Therefore, a combination of 
$m925s$, $Y$, and $J$ filters or $m925s$, $m975$, and $m1025$ filters
can determine 
whether such an object is a promising quasar candidate.
To securely identify the break, we need to reach $m925s \sim 22.3$ mag at 5$\sigma$, 
which is possible with a few hours of exposure time on a 2-m class telescope under nominal observing conditions.
 
\section{Summary} \label{summary}
We show the results from the medium-band observations of 6 high redshift quasars at 4.7 $\leq$ z $\leq$ 6.0 and 5 brown dwarfs. The SEDs of the quasars and brown dwarfs from the medium-band photometry 
show discrepancies that are much more significant than when using broad-band photometry, and their distant loci on various color-color diagrams can effectively distinguish the two populations.
The best-fit $\chi^2_{\nu}$ values of quasars and brown dwarfs are quite different even with information from only 3 medium-bands and SDSS photometry. These results from the medium-bands cannot be reproduced solely with SDSS photometry. 
Also, the $\chi^2$ fitting of the SEDs can estimate accurate z$_{\rm phot}$ within
an accuracy of $\Delta{\rm z}/(1+{\rm z}) \sim 0.012$
if we use the 5 medium-band filters discussed in this paper. 
Therefore, although previous quasar surveys are incapable of covering the redshift range between 5 and 6 due to the limitations of classic filter sets, our new technique using the medium-bands will be a powerful method for selecting high redshift quasars in this redshift gap. 
Two meter class telescopes, such as the McDonald 2.1-m telescope with SQUEAN, can conduct imaging follow-up observations of high redshift quasar candidates of $z$ $<$ 19.3 mag in these  3 medium-band filters within 15 minutes per target. For higher redshift quasars, larger telescopes are capable of carrying out the quasar survey at z $>$ 6 using longer bandpass medium-band filters with a high efficiency.


\acknowledgments
This work was supported by the National Research Foundation of Korea (NRF) grant, No. 2008-0060544,
funded by the Korean government (MSIP). 
M.H. acknowledges the support from Global PH.D Fellowship Program through the National Research Foundation of Korea (NRF) funded by the Ministry of Education (NRF-2013H1A2A1033110).
This paper includes data taken at The McDonald Observatory of The University of Texas at Austin.


\landscape
\begin{table}
\caption{Information of targets\label{tbl_sample}}
\centering
\begin{tabular}{cccccccc}
\toprule
Type&Name&$z$ (AB)&z$_{\rm spec}$&Integration Time$^{\rm a}$ (s)&Seeing$^{\rm b}$ ('')&z$_{\rm phot}$\\
\midrule

Quasar&IMS J032407.70+042613.3&19.15$\pm$0.06&4.712$\pm$0.039&         360,          360,          360,          180,          180&1.6, 1.6, 1.3, 1.3, 1.6&4.640$\pm$0.020\\    
Quasar&IMS J012247.33+121623.9&19.27$\pm$0.06&4.857$\pm$0.039$^{\rm c}$&        1200,         1200,         1200,          600,          600&3.7, 4.0, 4.0, 1.7, 1.7&4.870$\pm$0.045  \\  
Quasar&IMS J143704.82+070808.3&19.10$\pm$0.06&4.967$\pm$0.049&         360,          360,          360,          180,          180&1.6, 1.6, 1.6, 1.6, 1.6&4.810$\pm$0.055    \\
Quasar&IMS J222514.39+033012.6&19.47$\pm$0.10&5.350$\pm$0.049&         600,          600,          600,          600,          600&2.6, 2.3, 2.8, 1.7, 1.7&5.240$\pm$0.060    \\
Quasar&IMS J015533.28+041506.8&19.26$\pm$0.06&5.386$\pm$0.039&         360,          360,          360,          360,          600&2.4, 2.1, 1.9, 1.6, 1.4&5.330$\pm$0.070    \\
Quasar&IMS J014837.63+060020.1&19.29$\pm$0.06&5.993$\pm$0.013$^{\rm d}$&         720,          720,          720,          360,          360&2.1, 1.9, 1.6, 1.6, 1.3&5.950$\pm$0.025    \\
Brown Dwarf&IMS J232251.23+063830.1&18.99$\pm$0.06&...&         600,          600,          600,          600,          600&3.7, 3.7, 3.4, 1.7, 1.7&...\\
Brown Dwarf&IMS J005427.31+065127.0&19.30$\pm$0.09&...&        1200,         1200,         1200,          600,          600&2.8, 3.1, 2.6, 1.7, 1.7&...\\
Brown Dwarf&IMS J011601.62+052953.7&19.35$\pm$0.09&...&        1200,         1200,         1200,          600,          600&2.6, 3.7, 3.7, 1.7, 1.7&...\\
Brown Dwarf&IMS J033009.15+001404.3&19.17$\pm$0.11&...&         810,          720,          720,          360,          360&2.1, 2.1, 2.1, 1.9, 2.1&...\\
Brown Dwarf&IMS J031629.23+055729.2&19.37$\pm$0.06&...&         360,          360,          360,          180,          180&1.3, 1.3, 1.3, 1.3, 1.6&...\\
\bottomrule
\end{tabular}
\tabnote{
Note.
\\ $^{\rm a}$  Integration times of the $m675, m725, m775, m825,$ and $m875$ filters respectively, for the test observations
\\ $^{\rm b}$  Seeing values in the $m675, m725, m775, m825,$ and $m875$ filters respectively, for the test observations
\\ $^{\rm c}$  \citet{yi15} provide a more accurate redshift, z$_{\rm spec}=4.76$.
\\ $^{\rm d}$  \citet{jian15} provide a more accurate redshift, z$_{\rm spec}=5.923\pm0.003$.

}
\end{table}

\begin{table}
\tabcolsep=0.03cm
\centering
\caption{Photometric information of targets\label{tbl_mag}}
\begin{tabular}{cccccccccccccccccccccccccccccccccc}
\toprule
Name&$g$&$r$&$i$&$z$&$Y$&$J$&$H$&$K$&$m675$&$m725$&$m775$&$m825$&$m875$\\
\midrule
IMS J032407.70+042613.3&23.95$\pm$0.39&20.39$\pm$0.04&19.03$\pm$0.03&19.15$\pm$0.06&19.39$\pm$0.05&19.23$\pm$0.05&18.96$\pm$0.05&18.83$\pm$0.05&19.82$\pm$0.07&19.30$\pm$0.05&19.49$\pm$0.05&19.48$\pm$0.05&19.12$\pm$0.06\\
IMS J012247.33+121623.9&24.29$\pm$0.37&22.35$\pm$0.14&19.37$\pm$0.03&19.27$\pm$0.06&19.12$\pm$0.04&18.92$\pm$0.04&18.56$\pm$0.04&18.50$\pm$0.04&23.07$\pm$0.28&19.10$\pm$0.09&20.42$\pm$0.12&19.37$\pm$0.04&19.56$\pm$0.07\\
IMS J143704.82+070808.3&25.02$\pm$0.72&20.71$\pm$0.04&19.20$\pm$0.02&19.10$\pm$0.06&19.40$\pm$0.05&19.39$\pm$0.08&19.01$\pm$0.06&18.99$\pm$0.08&21.34$\pm$0.13&19.06$\pm$0.07&19.44$\pm$0.05&19.17$\pm$0.06&19.03$\pm$0.08\\
IMS J222514.39+033012.6&25.67$\pm$0.68&22.01$\pm$0.14&20.02$\pm$0.05&19.47$\pm$0.10&19.48$\pm$0.06&19.33$\pm$0.06&19.04$\pm$0.10&18.99$\pm$0.08&21.57$\pm$0.20&22.27$\pm$0.20&19.28$\pm$0.11&20.06$\pm$0.09&19.52$\pm$0.07\\
IMS J015533.28+041506.8&24.07$\pm$0.38&21.81$\pm$0.10&19.98$\pm$0.03&19.26$\pm$0.06&19.66$\pm$0.07&19.28$\pm$0.06&19.00$\pm$0.06&18.91$\pm$0.06&21.12$\pm$0.19&22.08$\pm$0.24&19.37$\pm$0.12&19.52$\pm$0.13&19.55$\pm$0.06\\
IMS J014837.63+060020.1&24.95$\pm$0.64&24.70$\pm$0.57&22.27$\pm$0.19&19.29$\pm$0.06&19.54$\pm$0.06&19.30$\pm$0.06&19.10$\pm$0.07&19.03$\pm$0.07&23.60$\pm$0.96&22.26$\pm$0.27&22.93$\pm$0.47&20.57$\pm$0.14&19.05$\pm$0.09\\
IMS J232251.23+063830.1&24.30$\pm$0.46&22.71$\pm$0.19&20.14$\pm$0.04&18.99$\pm$0.06&18.68$\pm$0.03&18.99$\pm$0.06&18.34$\pm$0.04&18.26$\pm$0.04&22.68$\pm$0.22&20.82$\pm$0.14&20.44$\pm$0.11&19.55$\pm$0.05&19.22$\pm$0.09\\
IMS J005427.31+065127.0&25.06$\pm$0.62&22.99$\pm$0.25&20.28$\pm$0.07&19.30$\pm$0.09&19.53$\pm$0.07&19.22$\pm$0.07&18.91$\pm$0.05&18.93$\pm$0.07&22.37$\pm$0.23&21.12$\pm$0.12&20.56$\pm$0.12&20.28$\pm$0.09&19.86$\pm$0.18\\
IMS J011601.62+052953.7&24.96$\pm$0.70&22.22$\pm$0.18&20.33$\pm$0.07&19.35$\pm$0.09&19.33$\pm$0.04&19.01$\pm$0.04&18.78$\pm$0.06&18.80$\pm$0.07&99.00$\pm$9.00&20.92$\pm$0.16&20.58$\pm$0.12&19.79$\pm$0.07&19.63$\pm$0.09\\
IMS J033009.15+001404.3&24.72$\pm$1.02&22.52$\pm$0.31&20.51$\pm$0.07&19.17$\pm$0.11&19.39$\pm$0.04&19.04$\pm$0.03&18.94$\pm$0.04&18.94$\pm$0.07&22.75$\pm$0.25&21.36$\pm$0.07&20.72$\pm$0.08&20.15$\pm$0.08&20.57$\pm$0.12\\
IMS J031629.23+055729.2&23.75$\pm$0.35&22.97$\pm$0.27&20.60$\pm$0.05&19.37$\pm$0.06&19.34$\pm$0.05&19.11$\pm$0.05&18.90$\pm$0.06&18.91$\pm$0.07&99.00$\pm$9.00&21.64$\pm$0.16&20.73$\pm$0.13&19.88$\pm$0.06&19.67$\pm$0.07\\
\bottomrule
\end{tabular}
\end{table}

\endlandscape

\end{document}